\documentclass[aps,prb,superscriptaddress,twocolumn]{revtex4-2}
\usepackage[utf8]{inputenc}
\usepackage{xcolor}
\usepackage{hyperref}
\usepackage{amsmath,bm}
\usepackage{graphicx}
\usepackage{mathtools}
\usepackage{mathrsfs}

\usepackage{graphicx}
\allowdisplaybreaks

\begin{document}

\title{Robust skyrmion nucleation and transport via composition gradient in rare earth-transition metal ferrimagnets}

\author{Pietro Diona}
\affiliation{Nanoscience, Scuola Normale Superiore, Piazza dei Cavalieri 7, Pisa, Italy}
\affiliation{Quantum Materials Theory, Italian Institute of Technology, Via Morego 30, Genoa, Italy}

\author{Luca Maranzana}
\affiliation{Quantum Materials Theory, Italian Institute of Technology, Via Morego 30, Genoa, Italy}
\affiliation{Department of Physics, University of Genoa, Via Dodecaneso 33, Genoa, Italy}
\author{Sergey Artyukhin}
\affiliation{Quantum Materials Theory, Genoa 16162, Italy}

\begin{abstract}
Magnetic skyrmions are typically nucleated via localized excitation methods, such as laser heating, spin–orbit torques, or engineered defects, and their controlled transport is often limited by the skyrmion Hall effect. Here, we demonstrate that spatial gradients in rare-earth--transition-metal alloys, achievable through patterned laser writing, enable both deterministic nucleation and controlled transport. The strategy leverages engineered compositional gradient to cross the magnetization compensation point, defining regions with opposite magnetization. An out-of-plane magnetic field selectively reverses the magnetization, generating a skyrmion. Crucially, the same compositional design enables a channel tuned to the spin angular momentum compensation point, where skyrmions are driven by spin–orbit torques at high velocity with vanishing skyrmion Hall effect. Local composition variations act as pinning sites, enabling threshold-controlled trapping and release.
\end{abstract}

\maketitle

\section{Introduction}
Magnetic domain walls, bubbles, and skyrmions represent key spin textures for racetrack memory technologies, proposed for the first time by Parkin et al. \cite{parkin2008magnetic}. Early research focused primarily on ferromagnetic devices, where domain wall nucleation and manipulation are achieved via local magnetic fields or spin-polarized current pulses \cite{ralph2008spin, brataas2012current, atkinson2003magnetic, metaxas2007creep, diona2022simulation}. With the growing demand for faster and more energy-efficient operation, increasing attention has been devoted to antiferromagnets \cite{baldrati2019mechanism, gomonay2016high, baltz2018antiferromagnetic, shiino2016antiferromagnetic, bilyk2026deterministic} and ferrimagnets \cite{manchon2019current, caretta2024domain, kim2022ferrimagnetic, siddiqui2018current} that enable ultrafast dynamics due to suppression of Walker breakdown \cite{zhang2023ferrimagnets}. Nucleation of spin textures remains a central challenge. Skyrmions and magnetic bubbles typically require a combination of interfacial Dzyaloshinskii–Moriya interaction (DMI), perpendicular magnetic anisotropy, and locally confined excitations \cite{powalla2022single}. 
These approaches generally rely on spatially localized perturbations---thermal, magnetic, or electronic---to overcome the topological energy barrier separating the skyrmionic texture from a trivial one \cite{kern2022deterministic}, and often require complex device geometries to counteract transverse skyrmion motion due to the skyrmion Hall effect \cite{jiang2017direct}.
Reported nucleation strategies include local laser heating \cite{olleros2022non, rottmayer2006heat, kryder2008heat, hono2018hamr, pan2009heat}, current-driven spin orbit torque (SOT) instabilities \cite{mallick2022current}, geometric constrictions \cite{zheng2019continuous}, local magnetic fields \cite{casiraghi2019individual}, magnetic defects \cite{de2023controlling}, and voltage-controlled anisotropy modulation \cite{kang2016voltage, olleros2025anatomy}. In contrast to previous approaches that rely on geometric patterning and localized excitations to both nucleate and control skyrmions \cite{he2023all}, here we propose a conceptually different strategy based on compositional gradients in rare-earth–transition-metal (RE–TM) ferrimagnets. This material platform offers high domain wall \cite{diona2026observation, sala2022asynchronous, sala2022ferrimagnetic, sala2023deterministic} and skyrmion velocities \cite{kim2017self, vakili2021computing}, while its amorphous metallic nature enables scalable deposition and straightforward integration into device architectures \cite{diona2026observation}. The RE concentration can be modified through laser-writer exposure \cite{riddiford2024grayscale}. We then exploit the tunability of the RE-TM composition by engineering a locally enhanced RE concentration.
Micromagnetic simulations show that this RE-TM composition gradient enables deterministic nucleation of bubbles or skyrmions under a uniform out-of-plane magnetic field, without any localized excitation. As a proof of concept, here we propose a device consisting of a cylindrical nucleation area with outgoing channels that are RE-rich and tuned to the spin angular momentum compensation point. In these regions, the reduced antiferromagnetic exchange interaction \cite{diona2026observation} lowers the exchange stiffness compared to the surrounding TM-rich concentration. This reduction enables the DMI to stabilize a skyrmion within an experimentally accessible regime. The circular geometry of the compensated RE-rich region promotes deterministic nucleation under a uniform magnetic field, while simultaneously providing conditions for high-velocity skyrmion motion and vanishing skyrmion Hall effect, enabling robust, strictly longitudinal motion under spin orbit torques. Furthermore, the compositional flexibility of RE-TM ferrimagnets enables local tuning of the magnetic properties along the transport channel. Such local modulation can be exploited to control skyrmion propagation and confinement through engineered energy barriers. More broadly, the results highlight how spatial control of the RE–TM composition can be used not only to tailor skyrmion nucleation, but also to define pathways for current-driven dynamics in compensated ferrimagnetic systems.

\begin{figure*}[t]
     \centering
         \includegraphics[width = 0.7\linewidth]{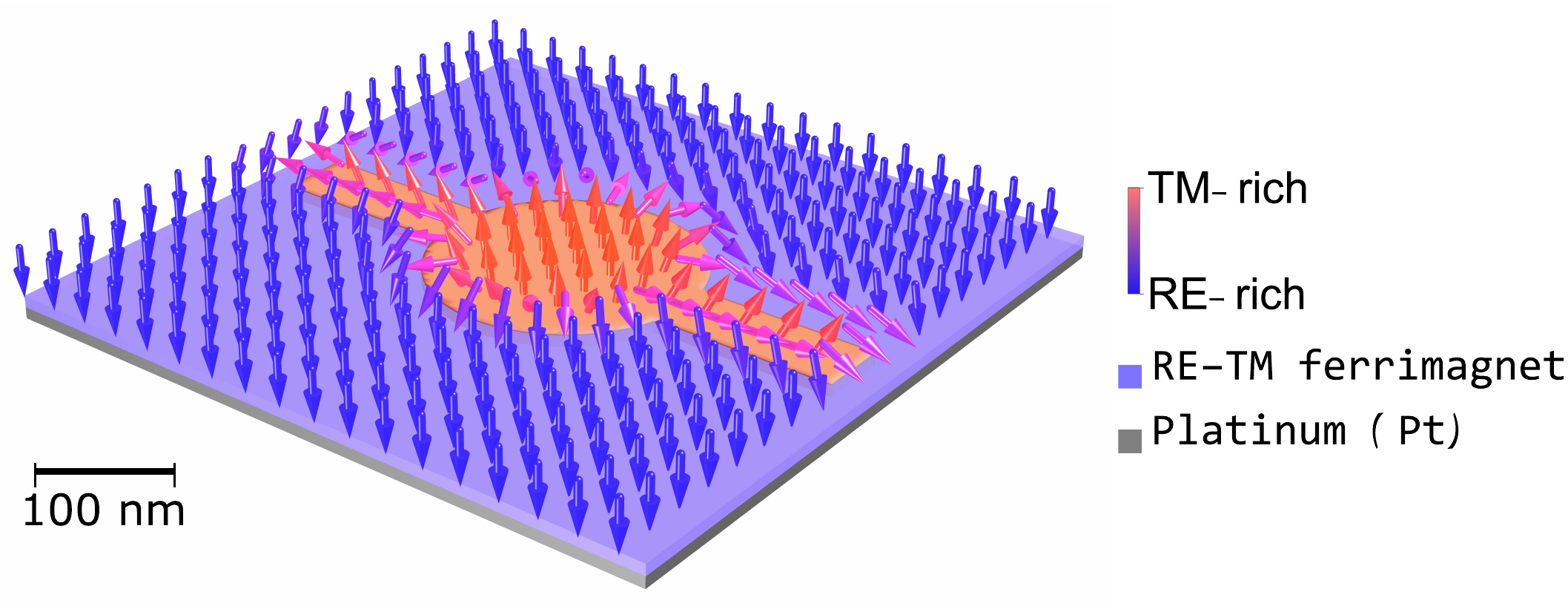}
    \caption{Three-dimensional representation of the engineered ferrimagnetic geometry and corresponding spin texture. The continuous RE-TM film (blue) hosts a disk-channel region (red) defined by a higher RE concentration. Arrows indicate the net spin orientation after the application of a uniform out-of-plane magnetic field, forming a skyrmion profile that extends across the entire patterned region. Spins point upward within the RE-rich geometry (disk and channel) and rotate toward a downward orientation in the surrounding area. The color scale (blue to red) reflects the local composition, from RE-rich to TM-rich regions. The gray layer is the Platinum layer which sources the SOT current.}
       \label{Fig1} 
\end{figure*}

\section{Main}
We consider a RE-TM ferrimagnetic film, where a disk-channel geometry is defined by spatial modulation of the RE concentration, as shown in Fig.~\ref{Fig1}. Laser exposition outside the red area leaves it RE-rich, tuned to the spin angular momentum compensation point. The system is initially annealed to a single domain state with a uniform out-of-plane Néel vector, which means the magnetization of the disk-channel region is opposite to that of the TM-rich surroundings. A uniform out-of-plane magnetic field applied along the magnetization of the latter drives the reversal of the order parameter of the disk-channel region. The exchange interaction favors the Néel vector to rotate smoothly across the boundary between the patterned region and the external material, resulting in the formation of a domain wall. In the presence of DMI, favoring a well-defined chirality, the system may cross the barrier, reaching a skyrmion configuration with finite topological charge Q. After the removal of the external field, the structure tends to contract due to the domain wall surface tension, but its topology prevents collapse, yielding a stable skyrmion, as shown in Fig.~\ref{Fig2}. The suppression of the spin angular momentum gives rise to zero skyrmion Hall effect. As a result, skyrmions driven by SOT propagate through the compensated channel with rectilinear trajectories and enhanced velocities \cite{kim2017self, vakili2021computing}. This enables deterministic transport from the nucleation site in the disk to the channel without deflections \cite{hirata2019vanishing, kolesnikov2018skyrmionium}. To achieve controlled manipulation of skyrmion motion, spatial variations of the RE concentration are introduced along the channel, resulting in a series of tunable pinning sites. These compositional gradients create local energy minima that can trap the skyrmion, while allowing depinning and propagation once a critical SOT current threshold is exceeded, as shown in Fig.~\ref{Fig3}. This mechanism provides a route to synchronize and control skyrmion dynamics under SOT-drive.

The proposed architecture therefore integrates skyrmion nucleation and guided transport within a single material platform. Importantly, nucleation is achieved using a spatially uniform magnetic field, without the need for localized excitations. At the same time, the spin angular momentum compensation ensures high speed and vanishing skyrmion Hall effect, while compositional engineering enables programmable pinning landscapes. This combination makes the design particularly attractive for experimental implementation and scalable spintronics applications.

\subsection{Micro-magnetic simulations}

\begin{figure*}[t]
     \centering
         \includegraphics[width = 0.95\linewidth]{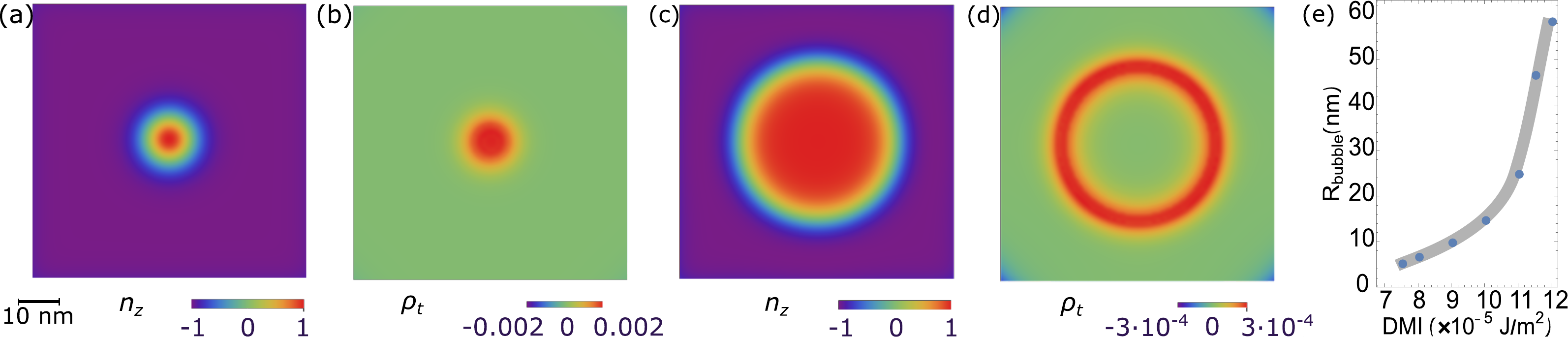}
    \caption{Antiferromagnetic order parameter $\hat{n}_z$ extracted from the relaxed micromagnetic configuration for: (a) chiral skyrmion with topological charge $Q = +1$; (c) skyrmion bubble. (b) and (d) show the corresponding topological charge densities. Panel (e) reports the equilibrium radius of the stabilized skyrmion bubble as a function of the DMI strength, highlighting the increase in the bubble size with increasing DMI. The ground states are obtained after applying an out-of-plane magnetic field ($+\hat{z}$) that reverses the magnetization of the Co-rich inner core, nucleating an intermediate magnetic bubble. Once the external field is removed, the system relaxes into a chiral skyrmion or a skyrmion bubble. In (a-d) the simulations are performed respectively with a realistic DMI strength of $7.5\cdot 10^{-5} \mathrm{J/m^2}$ and $1\cdot 10^{-4} \mathrm{J/m^2}$.}
       \label{Fig2} 
\end{figure*}

Micromagnetic simulations were performed using MuMax3, which was extended to accurately model ferrimagnetic systems \cite{vansteenkiste2014design, exl2014labonte}. We numerically solve the Landau–Lifshitz–Gilbert equation:

\begin{equation}
\label{LLG}
    \mathbf{\dot{m}} = -\frac{\gamma}{1+\alpha^2}\left[\mathbf{m}\times\mathbf{H_{\mathrm{eff}}} + \alpha\mathbf{m} \times\left(\mathbf{m}\times\mathbf{H_{\mathrm{eff}}}\right)\right] + \mathbf{\tau_{\mathrm{SOT}}},
\end{equation}

where $\gamma$ is the gyromagnetic ratio, $\alpha$ is the Gilbert damping parameter, $\tau_{\mathrm{SOT}}$ is the SOT originating from the platinum layer and reads as:

\begin{equation}
\boldsymbol{\tau}_{\mathrm{SOT}}
=
-\gamma H_{\mathrm{DL}}\,
\mathbf{m}\times\left(\mathbf{m}\times\boldsymbol{\sigma}\right)
+\gamma H_{\mathrm{FL}}\,
\mathbf{m}\times\boldsymbol{\sigma},
\end{equation}
where $H_{\mathrm{DL}}$ and $H_{\mathrm{FL}}$ denote the damping-like and field-like effective fields, respectively, and $\boldsymbol{\sigma}$ is the spin polarization direction. Throughout this work only the damping-like torque is considered ($H_{\mathrm{FL}}=0$), consistent with Pt/RE--TM ferrimagnetic heterostructures \cite{diona2026nucleation, diona2026observation, sala2022asynchronous, sala2023deterministic, sala2022ferrimagnetic}. The damping-like field is $H_{\mathrm{DL}} = 
\frac{\hbar\,\theta_{\mathrm{SH}}\,J}
{2e\,\mu_0 M_s t}.$

In Eq.~\ref{LLG}, $\mathbf{H_{\mathrm{eff}}} = -\frac{1}{M_{\mathrm{s}}}\frac{\delta\mathcal{H}}{\delta\mathbf{m}}$ is the effective field where $\mathcal{H}$ is the Hamiltonian of the system: 
\begin{widetext}
\begin{equation}
    \mathcal{H} = A\left(\nabla \mathbf{m}\right)^2 - K_u\left(\hat{z}\cdot\mathbf{m}\right)^2 + D_0\left[m_z\left(\nabla\cdot\mathbf{m}\right)-\left(\mathbf{m}\cdot\nabla\right)m_z\right] - M_s\mathbf{m}\cdot\mathbf{B_{ext}},
\end{equation}
\end{widetext}
where $A$ is the exchange interaction, $K_u$ is the perpendicular magnetic anisotropy, $D_0$ is the DMI constant, and $\mathbf{B_{ext}}$ is the external magnetic field driving the nucleation. Model parameters are based on the experimentally measured values for GdFeCo (Tab. \ref{param}) as reported in \cite{diona2026observation, sala2022asynchronous, diona2026variable}. The lattice spacing of $d = 1$~nm was used in the simulations. The simulated system is discretized on a $(768 \times 256 \times 1)$ computational grid with a cell size of $(1\times1\times1)\,\mathrm{nm^3}$, corresponding to a simulated area of $(768\times256)\,\mathrm{nm^2}$. Since the magnetization is assumed to remain uniform across the film thickness, only one computational cell is used along the out-of-plane direction, following the standard thin-film methodology adopted in micromagnetic simulations.
To investigate the dependence of  the reported results on the damping parameter, the nucleation simulations were repeated using the experimentally realistic value $\alpha=0.01$ \cite{diona2026nucleation, diona2026observation, bainsla2022ultrathin, kim2018low}. We find that a lower damping increases the relaxation time, requiring a longer out-of-plane magnetic-field pulse (20 ns instead of 2 ns) to recover the same nucleation mechanism and final magnetic configuration. The simulations reported in this work employ $\alpha=0.1$ to reduce the computational cost.
The simulated structure consists of two regions with distinct ferrimagnetic compositions. Outside the disk-channel region, the parameters correspond to a Co-rich--GdFeCo alloy above the magnetization compensation point \cite{diona2026observation, sala2022asynchronous}. Inside the disk-channel region, the parameters reproduce a Gd-rich composition tuned to the spin angular momentum compensation point \cite{diona2026observation, sala2022asynchronous}. This concentration gradient reflects the experimentally achievable profile of RE and TM concentrations in amorphous ferrimagnetic alloys, through  non-uniform laser exposure \cite{riddiford2024grayscale}. The DMI constant is varied within a realistic range of $7 \cdot 10^{-5}$ to $1.3 \cdot 10^{-4}$ $\mathrm{J/m^2}$, in agreement with previous experimental reports \cite{morshed2021tuning, kim2019bulk}. This interval corresponds to values achievable experimentally by adjusting the relative thickness of the GdFeCo layer with respect to the underlying Pt layer. The magnetic ground state obtained after applying and removing an out-of-plane uniform magnetic field depends sensitively on the DMI strength. For DMI values between $7.5 \cdot 10^{-5}$ and $1.2 \cdot 10^{-4}$ $\mathrm{J/m^2}$, the system relaxes into a stable skyrmion or skyrmion bubble, as shown in Fig.~\ref{Fig2}(a,b,c,d).
The diameter of the skyrmion bubble increases monotonically with the DMI strength, as shown in Fig.~\ref{Fig2}(e).
For DMI values greater than $1.3 \cdot 10^{-4}$ $\mathrm{J/m^2}$, which are generally difficult to achieve experimentally in these systems, the ground state transits into a spin spiral configuration.

This simulation framework therefore captures the transition between skyrmions and skyrmion bubbles, arising from experimentally realistic material and DMI strengths.

\begin{table}[!b]
 \caption{Simulation parameters}
    \begin{tabular}{@{}lll@{}} 
    \hline
    Parameters & Outer region &  Inner core\\
    \hline\hline
    $A$  & $\mathrm{4 \, pJ/m}$ & $\mathrm{1.4 \, pJ/m}$\\
    \hline
    $M_{\mathrm{A}}$ & $\mathrm{3\cdot 10^{5} \,A/m}$ & $\mathrm{4.4\cdot 10^{5} \,A/m}$\\
     \hline
    $M_{\mathrm{B}}$ & $\mathrm{4\cdot 10^{5} \,A/m}$ & $\mathrm{4\cdot 10^{5} \,A/m}$\\
    \hline
    $\alpha$ & 0.1 & 0.1\\
    \hline
    $K_u$ & $8\cdot 10^3\,\mathrm{J/m}^3$ &  $1.2\cdot 10^4\,\mathrm{J/m}^3$\\
     \hline
     $\mathrm{D_0}$ & $7.5 \cdot 10^{-5} \, \mathrm{J/m^2}$ & $\mathrm{7.5 \cdot 10^{-5} \, J/m^2}$\\
     \hline
     $d$ & 1 nm & 1 nm\\
     \hline
    $\theta_{\mathrm{SH}}$  & 0.13 & 0.13 \\
      \hline
     $\mathrm{Grid \, size}$ & 768 x 256 x 1 & -\\
      \hline
     $\mathrm{Cell\,size}$ & 1 x 1 x 1 $\mathrm{nm}^3$ & - \\
      \hline 
    \end{tabular}
    \label{param}
\end{table}

\begin{figure*}[!t]
    \centering
    \includegraphics[width=\textwidth]{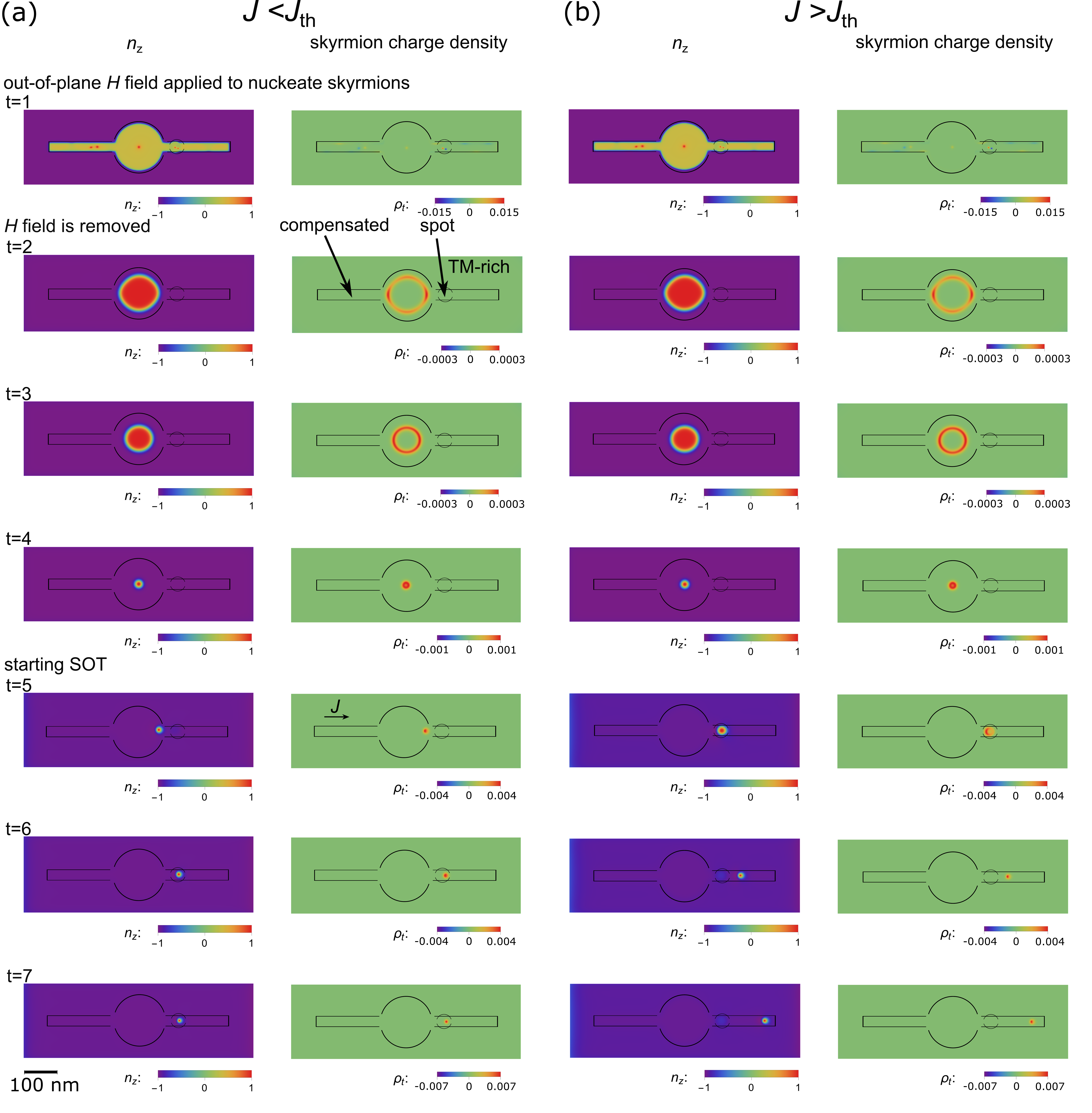}
    \caption{Micromagnetic simulations of skyrmion nucleation, zero skyrmion Hall effect transport, and pinning under spin–orbit torque (SOT) for two current regimes: (a) sub-threshold ($J < J_{\mathrm{th}}$) and (b) above-threshold ($J > J_{\mathrm{th}}$). A skyrmion is deterministically nucleated at the center of the RE-rich region under a uniform out-of-plane magnetic field and subsequently driven along the channel by SOT. The skyrmion moves with zero skyrmion Hall effect, because of spin angular momentum compensation point. In the sub-threshold regime (a), the skyrmion is pinned by a localized RE-rich pinning site. The reported depinning threshold refers exclusively to the engineered pinning potential. Intrinsic pinning arising from the amorphous microstructure is not included in the simulations, since it strongly depends on the growth conditions and would increase the depinning current without modifying the trapping mechanism. In contrast, for current densities exceeding the threshold (b), the applied SOT overcomes the pinning potential, leading to propagation along the channel.}
    \label{Fig3}
\end{figure*}

The Pt underlayer beneath the ferrimagnet not only provides interfacial DMI stabilizing the chirality of the magnetic texture, but it is also essential for manipulating the spin textures via SOT \cite{diona2026observation, diona2022simulation, sala2022ferrimagnetic, diona2026nucleation}.
We now turn to simulate the current-driven dynamics of the skyrmion within the engineered disk-channel geometry. As shown in Fig.~\ref{Fig3}, we compare two regimes corresponding to sub-threshold and above-threshold SOT current densities. The threshold current density required to depin the skyrmion from the engineered parking spot is found to be approximately $J_{\rm th}=3.5\times10^{10}\,\mathrm{A/m^2}$ ($J_{\rm th}=2.5\times10^{10}\,\mathrm{A/m^2}$) for $\alpha=0.1$ ($\alpha=0.01$). Following the deterministic nucleation at the center of the RE-rich region, the skyrmion is driven along the channel by the applied SOT. The current is modeled as a rectangular pulse with instantaneous rise and fall. Since the SOT is employed only to transport the nucleated skyrmion, whereas nucleation itself is driven by the external magnetic field, the specific temporal profile of the current pulse is not expected to affect the nucleation mechanism. Edge-induced distortions of the current distribution are neglected because the nucleation region and the parking spot are located inside the nanowire, where the current density can be considered uniform, as is experimentally observed in \cite{diona2026nucleation, diona2026observation}. Due to the spin angular momentum compensation of the disk-channel structure, the skyrmion motion remains longitudinal, without exhibiting skyrmion Hall effect. In the sub-threshold regime ($J < J_{\mathrm{th}}$), the skyrmion propagates until it reaches a localized RE-rich region acting as a pinning site. This “parking spot” traps the skyrmion due to the lower antiferromagnetic exchange interaction and anisotropy \cite{diona2026observation}. In contrast, for current densities exceeding the threshold ($J > J_{\mathrm{th}}$), the skyrmion overcomes the pinning potential, leading to depinning and propagation along the channel. These results demonstrate deterministic skyrmion transport with threshold-controlled trapping and release, enabled by compositional engineering and compensated dynamics.

\section{Conclusions}
We have introduced a compositional design strategy for the deterministic nucleation and control of skyrmions in RE-TM ferrimagnets. Through micromagnetic simulations, we demonstrate that spatial modulation of the RE-TM composition enables selective magnetization reversal within a RE-rich core under a uniform out-of-plane magnetic field, without the need for localized excitation. This mechanism exploits the reduced antiferromagnetic exchange interaction to stabilize skyrmions within an experimentally accessible DMI regime. Beyond nucleation, the same design defines confined channels tuned to the spin angular momentum compensation point, enabling current-driven skyrmion motion with vanishing skyrmion Hall effect. As a consequence, skyrmion propagates longitudinally under spin–orbit torque current. Furthermore, local modulation of the RE concentration provides a route to engineer pinning sites with well-defined depinning thresholds, allowing controlled trapping and release of skyrmions. Our results establish a unified framework in which compositional engineering replaces geometric complexity and localized stimuli, enabling deterministic nucleation, Hall-free transport, and programmable manipulation of skyrmions using uniform magnetic field and current. This approach offers a scalable and experimentally viable pathway toward skyrmion-based devices and functional spintronic architectures.

\section*{Acknowledgements}
P.D., L.M. acknowledge funding from IIT Genova. 

\section*{Conflict of Interest} 
The authors declare no conflict of interest.


\newpage
\bibliography{bibliographyMain}

\end{document}